\documentclass[twocolumn,showpacs,aps,prl,superscriptaddress,preprintnumbers,amsmath,amssymb]{revtex4}

\usepackage{graphicx}
\usepackage{dcolumn}
\usepackage{bm}
\usepackage{amsmath}
\usepackage{epsfig}

\RequirePackage{xspace}

\usepackage{relsize}
\def\babar{\mbox{\slshape B\kern-0.1em{\smaller A}\kern-0.1em
    B\kern-0.1em{\smaller A\kern-0.2em R}}}

\def\epem       {\ensuremath{e^+e^-}\xspace}

\def\qqbar {\ensuremath{q\overline q}\xspace}
\def\u     {\ensuremath{u}\xspace}
\def\ubar  {\ensuremath{\overline u}\xspace}

\def\d     {\ensuremath{d}\xspace}

\def\b     {\ensuremath{b}\xspace}

\def\piz   {\ensuremath{\pi^0}\xspace}

\def\Kbar  {\kern 0.2em\overline{\kern -0.2em K}{}\xspace}

\def\Kz    {\ensuremath{K^0}\xspace}
\def\Kzb   {\ensuremath{\Kbar^0}\xspace}
\def\KzKzb {\ensuremath{\Kz \kern -0.16em \Kzb}\xspace}
\def\Kp    {\ensuremath{K^+}\xspace}
\def\Km    {\ensuremath{K^-}\xspace}

\def\KpKm  {\ensuremath{\Kp \kern -0.16em \Km}\xspace}

\def\Dbar    {\kern 0.2em\overline{\kern -0.2em D}{}\xspace}

\def\Dz      {\ensuremath{D^0}\xspace}
\def\Dzb     {\ensuremath{\Dbar^0}\xspace}
\def\DzDzb   {\ensuremath{\Dz {\kern -0.16em \Dzb}}\xspace}
\def\Dp      {\ensuremath{D^+}\xspace}
\def\Dm      {\ensuremath{D^-}\xspace}

\def\DpDm    {\ensuremath{\Dp {\kern -0.16em \Dm}}\xspace}

\def\B       {\ensuremath{B}\xspace}
\def\Bbar    {\kern 0.18em\overline{\kern -0.18em B}{}\xspace}

\def\BB      {\ensuremath{B\Bbar}\xspace} 
\def\Bz      {\ensuremath{B^0}\xspace}
\def\Bzb     {\ensuremath{\Bbar^0}\xspace}
\def\BzBzb   {\ensuremath{\Bz {\kern -0.16em \Bzb}}\xspace}
\def\Bu      {\ensuremath{B^+}\xspace}
\def\Bub     {\ensuremath{B^-}\xspace}
\def\Bp      {\ensuremath{\Bu}\xspace}

\def\Bpm     {\ensuremath{B^\pm}\xspace}

\def\BpBm    {\ensuremath{\Bu {\kern -0.16em \Bub}}\xspace}

\def\BorBbar    {\kern 0.18em\optbar{\kern -0.18em B}{}\xspace}
\def\DorDbar    {\kern 0.18em\optbar{\kern -0.18em D}{}\xspace}
\def\KorKbar    {\kern 0.18em\optbar{\kern -0.18em K}{}\xspace}

\mathchardef\Upsilon="7107
\def\Y#1S{\ensuremath{\Upsilon{(#1S)}}\xspace}

\def\FourS {\Y4S}

\mathchardef\Deltares="7101
\mathchardef\Xi="7104
\mathchardef\Lambda="7103
\mathchardef\Sigma="7106
\mathchardef\Omega="710A

\def\Deltabar{\kern 0.25em\overline{\kern -0.25em \Deltares}{}\xspace}
\def\Lbar{\kern 0.2em\overline{\kern -0.2em\Lambda\kern 0.05em}\kern-0.05em{}\xspace}
\def\Sigbar{\kern 0.2em\overline{\kern -0.2em \Sigma}{}\xspace}
\def\Xibar{\kern 0.2em\overline{\kern -0.2em \Xi}{}\xspace}
\def\Obar{\kern 0.2em\overline{\kern -0.2em \Omega}{}\xspace}
\def\Nbar{\kern 0.2em\overline{\kern -0.2em N}{}\xspace}
\def\Xb{\kern 0.2em\overline{\kern -0.2em X}{}\xspace}

\def\upsbb   {\ensuremath{\FourS \to \BB}\xspace}

\def\mes        {\mbox{$m_{\rm ES}$}\xspace}

\def\DeltaE     {\mbox{$\Delta E$}\xspace}

\newcommand{\tev}{\ensuremath{\mathrm{\,Te\kern -0.1em V}}\xspace}
\newcommand{\gev}{\ensuremath{\mathrm{\,Ge\kern -0.1em V}}\xspace}
\newcommand{\mev}{\ensuremath{\mathrm{\,Me\kern -0.1em V}}\xspace}
\newcommand{\kev}{\ensuremath{\mathrm{\,ke\kern -0.1em V}}\xspace}
\newcommand{\ev}{\ensuremath{\mathrm{\,e\kern -0.1em V}}\xspace}
\newcommand{\gevc}{\ensuremath{{\mathrm{\,Ge\kern -0.1em V\!/}c}}\xspace}
\newcommand{\mevc}{\ensuremath{{\mathrm{\,Me\kern -0.1em V\!/}c}}\xspace}
\newcommand{\gevcc}{\ensuremath{{\mathrm{\,Ge\kern -0.1em V\!/}c^2}}\xspace}
\newcommand{\mevcc}{\ensuremath{{\mathrm{\,Me\kern -0.1em V\!/}c^2}}\xspace}

\def\mus  {\ensuremath{\rm \,\mus}\xspace}

\def\ps   {\ensuremath{\rm \,ps}\xspace}

\def\mus        {\ensuremath{\,\mu{\rm s}}\xspace}    
\def\ps         {\ensuremath{{\rm \,ps}}\xspace}  

\def\to                 {\ensuremath{\rightarrow}\xspace}

\newcommand{\stat}{\ensuremath{\mathrm{(stat)}}\xspace}
\newcommand{\syst}{\ensuremath{\mathrm{(syst)}}\xspace}

\def\pep2{PEP-II}

\def\gsim{{~\raise.15em\hbox{$>$}\kern-.85em
          \lower.35em\hbox{$\sim$}~}\xspace}
\def\lsim{{~\raise.15em\hbox{$<$}\kern-.85em
          \lower.35em\hbox{$\sim$}~}\xspace}

\def\CP                {\ensuremath{C\!P}\xspace}
\def\C       {\ensuremath{C}\xspace}

\def\deltaz{\ensuremath{{\rm \Delta}z}\xspace}
\def\deltat{\ensuremath{{\rm \Delta}t}\xspace}
\def\deltamd{\ensuremath{{\rm \Delta}m_d}\xspace}

\xspace

\newcommand{\epjBase}        {Eur.\ Phys.\ Jour.\xspace}
\newcommand{\jprlBase}       {Phys.\ Rev.\ Lett.\xspace}
\newcommand{\jprBase}        {Phys.\ Rev.\xspace}
\newcommand{\jplBase}        {Phys.\ Lett.\xspace}
\newcommand{\nimBaseA}       {Nucl.\ Instr.\ Methods Phys.\ Res., Sect.\ A\xspace}

\newcommand{\epjc}      [1]  {\epjBase\ C~{\bf #1}}

\newcommand{\nima}      [1]  {\nimBaseA~{\bf #1}}

\newcommand{\plb}       [1]  {\jplBase\ B~{\bf #1}}

\newcommand{\jprl}      [1]  {\jprlBase\ {\bf #1}}
\newcommand{\jprd}      [1]  {\jprBase\ D~{\bf #1}}

\newcommand{\progtp}    [1]  {{Prog.\ Theor.\ Phys.\ {\bf #1}}}

\def\jetset74   {\mbox{\tt Jetset \hspace{-0.5em}7.\hspace{-0.2em}4}\xspace}

\newcommand{\BABARPubYear}    {05}
\newcommand{\BABARPubNumber}  {007}

\newcommand{\SLACPubNumber} {11081}
\newcommand{\LANLNumber} {0503049}
\def\Bztorhoprhom {\ensuremath{\Bz (\Bzb) \to \rho^+ \rho^- }\xspace}

\def\clong   {\ensuremath{ C_{L} }}
\def\slong   {\ensuremath{ S_{L} }}
\def\ctran   {\ensuremath{ C_{T} }}
\def\stran   {\ensuremath{ S_{T} }}
\def\fL      {\ensuremath{ f_L }}
\def\ptrue   { \fL }

\def\coshel  {\ensuremath{ \cos\theta_{i} }}
\def\mv      {\ensuremath{ m_{\pi^\pm \pi^0 }}}

\def\nno     {\ensuremath{\cal{N}}}

\def\rhop {\ensuremath{ \rho^+ } }
\def\rhom {\ensuremath{ \rho^- } }
\def\rhoz {\ensuremath{ \rho^0 } }

\def\figurebox#1#2#3{%
    \def\arg{#3}%
    \ifx\arg\empty
    {\hfill\vbox{\hsize#2\hrule\hbox to #2{\vrule\hfill\vbox to #1{\hsize#2\vfill}\vrule}\hrule}\hfill}%
    \else
    {\hfill\epsfbox{#3}\hfill}%
    \fi}

\begin{document}

\preprint{\babar-PUB-\BABARPubYear/\BABARPubNumber}
\preprint{SLAC-PUB-\SLACPubNumber}
\begin{flushleft}
hep-ex/\LANLNumber\\[10mm]
\end{flushleft}

\title{
{\large \bf Improved Measurement of the CKM Angle {\boldmath $\alpha$} Using
{\boldmath \Bztorhoprhom} Decays. }}

%
\author{B.~Aubert}
\author{R.~Barate}
\author{D.~Boutigny}
\author{F.~Couderc}
\author{Y.~Karyotakis}
\author{J.~P.~Lees}
\author{V.~Poireau}
\author{V.~Tisserand}
\author{A.~Zghiche}
\affiliation{Laboratoire de Physique des Particules, F-74941 Annecy-le-Vieux, France }
\author{E.~Grauges}
\affiliation{IFAE, Universitat Autonoma de Barcelona, E-08193 Bellaterra, Barcelona, Spain }
% ...
\collaboration{The \babar\ Collaboration}
\noaffiliation

\date{\today}

\begin{abstract}

%
%
We present results from an analysis of \Bztorhoprhom\ using 232
million \upsbb decays collected with the \babar\ detector at the
\pep2\ asymmetric-energy $B$ Factory at SLAC. 
We measure the longitudinal polarization fraction $\ptrue = 0.978
\pm 0.014 \stat \,^{+0.021}_{-0.029} \syst$ and the \CP-violating
parameters ${\slong} = -0.33 \pm 0.24 \stat ^{+0.08}_{-0.14} \syst$
and $\clong = -0.03\pm 0.18 \stat \pm 0.09 \syst$. Using an isospin
analysis of $B\rightarrow \rho\rho$ decays we determine the 
unitarity triangle parameter $\alpha$. 
The solution compatible with the Standard Model is $\alpha = 
(100 \pm 13)^\circ$.

\end{abstract}


\maketitle

In the Standard Model, \CP-violating effects in the \B-meson 
system arise from a single phase in the 
Cabibbo-Kobayashi-Maskawa (CKM) quark-mixing matrix~\cite{CKM}.
Interference between direct decay and
decay after $\Bz\Bzb$ mixing in \Bztorhoprhom results 
in a time-dependent decay-rate asymmetry that is sensitive 
to the angle $\alpha \equiv 
\arg\left[-V_{td}^{}V_{tb}^{*}/V_{ud}^{}V_{ub}^{*}\right]$ 
in the unitarity triangle of the CKM matrix . 
This decay proceeds mainly through a $\b \to \u\ubar \d$ tree diagram.
The presence of penguin loop contributions introduces additional
phases that shift the experimentally measurable parameter
$\alpha_{\mathrm{eff}}$ away from the value of $\alpha$. 
However, measurements of the
$\Bp \to \rho^+\rho^0$ branching fraction and the upper limit for $\Bz
\to \rho^0 \rho^0$ \cite{recentrhorho,PRLrho0rho0} show that the penguin
contribution in $\B \to \rho \rho$ is small with respect to the leading 
tree diagram, and $\delta\alpha_{\rho\rho} = \alpha_{\mathrm{eff}} - \alpha$
is constrained at $\pm 11^\circ$ at $1\sigma$~\cite{PRLrho0rho0}.
This Letter presents an update of the time-dependent analysis of 
\Bztorhoprhom and measurement of the CKM angle $\alpha$ reported 
in \cite{ref:us}.

The \CP analysis of $B$ decays to $\rho^+\rho^-$ is complicated 
by the presence of a mode with longitudinal polarization and 
two with transverse polarizations.  The longitudinal mode is \CP 
even, while the transverse modes contain \CP-even and \CP-odd states.
 Empirically, the decay is observed to
be dominated by the longitudinal polarization \cite{ref:us}, with a fraction 
$\ptrue$ defined by the fraction of the helicity zero state in the decay. The angular 
distribution is
\begin{eqnarray}
&&\frac{d^2\Gamma}{\Gamma d\cos\theta_1 d\cos\theta_2}= \label{eqn:one} \\  \nonumber
&& \frac{9}{4}\left[f_L \cos^2\theta_1 \cos^2\theta_2 + \frac{1}{4}(1-\ptrue) \sin^2\theta_1 \sin^2\theta_2 \right]
\end{eqnarray}
where $\theta_{i=1,2}$ is the angle between the \piz momentum 
and the direction opposite the $B^0$ in the $\rho$ rest 
frame, and we have integrated over the angle between 
the $\rho$ decay planes.

The analysis reported here is improved over our earlier publication 
\cite{ref:us} by a change in selection requirements resulting in 
an increased signal efficiency; introduction of a signal time 
dependence that accounts for possible misreconstruction; and use of 
a more detailed background model.  
This measurement uses 232 million \upsbb decays collected
with the \babar ~\cite{ref:babar} detector at the \pep2\
asymmetric-energy $B$ Factory at SLAC.

We reconstruct \Bztorhoprhom candidates ($B_{\rm rec}$) from combinations of
two charged tracks and two \piz candidates. We require that both tracks have
particle identification information inconsistent with the electron, kaon, and
proton hypotheses. The \piz candidates
are formed from pairs of photons each of which has a measured energy 
greater than
$50~\mev$. The reconstructed \piz\ mass must satisfy
$0.10 < m_{\gamma\gamma} < 0.16~\gevcc$. The mass of the $\rho$ candidates
must satisfy $0.5 < \mv < 1.0~\gevcc$. When multiple \B candidates can
be formed, we select the one that minimizes the sum of $( m_{\gamma\gamma} - m_{\piz} )^2$
where $m_{\piz}$ is the true \piz mass.  If more than one 
candidate has the same \piz\ mesons, we select one at random.

Combinatorial backgrounds dominate near $|\coshel|=1$,
and backgrounds from $B$ decays tend to concentrate at
negative values of $\coshel$. We reduce these backgrounds with the
requirement $-0.90 < \coshel < 0.98$.

Continuum $\epem \to \qqbar$ ($q = u,d,s,c$) events are the dominant
background.  This background is reduced by requiring 
that $|\cos\B_{TR}|<0.8$, where $\B_{TR}$ is the angle between the $B$ 
thrust axis and that of the rest of the event, ROE.
The thrust axis of the \B is the direction 
which maximizes the longitudinal momenta of the particles in the \B\ candidate. To 
distinguish signal from continuum we use a neural network
(\nno) to combine ten discriminating variables: the event shape variables
that are used in the Fisher discriminant in Ref~\cite{pipiBabar};
the cosine of
the angle between the direction of the \B and the collision axis ($z$) in the
\epem center-of-mass (CM) frame; the cosine of the angle between the \B thrust
axis and the $z$ axis, $|\cos\B_{TR}|$;
the decay angle of each \piz (defined in
analogy to the $\rho$ decay angle, $\theta_i$); and the sum of transverse
momenta in the ROE relative to the $z$ axis. 

Signal events are identified kinematically using two variables, the
difference \DeltaE between the CM energy of the \B candidate and
$\sqrt{s}/2$, and the beam-energy-substituted mass
 $\mes = \sqrt{(s/2 + {\mathbf {p}}_i\cdot {\mathbf {p}}_B)^2/E_i^2- {\mathbf {p}}_B^2}$,
where $\sqrt{s}$ is the total CM energy. The \B momentum
${\mathbf {p}_B}$ and four-momentum of the initial state $(E_i, {\mathbf
{p}_i})$ are defined in the laboratory frame. We accept candidates that
satisfy $5.23 < \mes <5.29~\gevcc$ and $-0.12<\DeltaE<0.15~\gev$. The
asymmetric \DeltaE selection reduces background from higher-multiplicity
\B decays.

To study the time-dependent asymmetry one needs to measure the proper-time 
difference, \deltat, between the two \B\ decays in the event, and
to determine the flavor of the other \B\ meson ($B_{\rm tag}$).
We calculate \deltat from the measured separation
\deltaz between the $B_{\rm rec}$ and $B_{\rm tag}$ decay
vertices~\cite{prdsin2b}. We determine the $B_{\rm rec}$ vertex
from the two charged-pion tracks in its decay. The $B_{\rm tag}$
decay vertex is obtained by fitting the other tracks in the event,
with constraints from the $B_{\rm rec}$ momentum and the
beam-spot location. The RMS resolution on
$\deltat$ is 1.1 \ps. We only use events that satisfy 
$|\deltat|<20 \, \ps$ and for which the error on \deltat 
less than $2.5 \, \ps$.
The flavor of the $B_{\rm tag}$ meson is determined with a multivariate
technique~\cite{pipiBabar} that has a total effective tagging efficiency of
$(29.9\pm 0.5)$\%.  

Signal candidates may pass the selection requirement even 
if one or more of the pions assigned to the $\rho^+\rho^-$ 
state belongs to the other $B$ in the event. These 
self-cross-feed (SCF) candidates constitute 50\% (26\%) of the 
accepted signal for $\fL=1$ ($\fL=0$).   The majority of SCF 
events have both charged pions from the $\rho^+\rho^-$ final state,
and unbiased \CP information (correct-track SCF).  There 
is a SCF component (14\% of the signal) where at least one track 
in $B_{\rm rec}$ is from the rest of the event.  These wrong 
track events have biased \CP information, and are treated 
separately for the \CP result. The probability density 
function (PDF) describing wrong
track events is used only in determining the signal yield 
and polarization. A systematic error is assigned to the \CP 
results from this type of signal event.

We obtain a sample of 68703 events that enter a maximum-likelihood fit.  
These events are dominated by
backgrounds: roughly $92$\% from \qqbar and $7$\% from \BB events.  
The remaining 1\% of events is signal.
We distinguish the following candidate types:
 (i) correctly reconstructed signal;
 (ii) SCF signal, split into correct and wrong track parts;
 (iii) charm $\Bpm$ background ($b\to c$);
 (iv) charm $\Bz$ background ($b\to c$);
 (v) charmless $B$ backgrounds; and
 (vi) continuum background. 
The dominant charmless backgrounds are \B\ decays to 
$\rho\pi$, $(a_1 \pi)^\pm$, $(a_1 \pi)^0$, and 
longitudinally polarized $a_1\rho$ final states. 
For these decays we use the inclusive branching fractions 
(in units of $10^{-6}$), $34 \pm 4$~\cite{rhopi}, $42\pm 42$, 
$42\pm 6$~\cite{aonepi} and  $100\pm 100$, respectively. 
The corresponding expected number of events in the sample are
$82 \pm 13$, $87\pm 87$, $65\pm 9$, and 
$202\pm 202$. We also account for contributions from higher  
kaon resonances ($112 \pm 112$ events) and 
 $\rho^+\rho^0$ ($82\pm19$ events). 
In addition we expect $2551 \pm 510$ ($1316 \pm 263$) charged (neutral) \B\ decays to final states containing
charm mesons. The \B-background decays are included as 
separate components in the fit.

Each candidate is described with the eight $B_{\rm rec}$ kinematic
variables: \mes, \DeltaE, the \mv\ and \coshel\ values of the two $\rho$
mesons, \deltat, and \nno.  For each fit component,
we construct a PDF that is the product of PDFs
for these variables, neglecting correlations.  This introduces 
a fit bias that is corrected with the use of Monte Carlo (MC) simulation.
The continuum-background yield and its PDF parameters for \mes, 
\DeltaE, \coshel, and \nno\ are floated in the
fit to data. The continuum \mv\ distribution is described by a 
Breit-Wigner and polynomial shape, and is derived from \mes\ and \DeltaE\ data 
sidebands. For all other fit components the PDFs
 are extracted from high-statistics MC 
samples. The \coshel\ distributions for the background are 
described by a non-parametric (NP) PDF derived from the MC samples, as the detector
acceptance and selection modify the known vector-meson decay 
distribution. The true signal distribution is given by Eq.~\ref{eqn:one} multiplied 
by an acceptance function determined from signal MC samples, whereas SCF signal
is modeled using NP PDFs.

The signal decay-rate distribution for both polarizations
$f_+ (f_-)$ for $B_{\rm tag}$= \Bz (\Bzb) is given by
\begin{eqnarray*}
f_{\pm}(\deltat) = \frac{e^{-\left|\deltat\right|/\tau}}{4\tau} [1
\pm S\sin(\deltamd\deltat) \mp \C\cos(\deltamd\deltat)]\,,  \nonumber
\end{eqnarray*}
where $\tau$ is the mean \Bz lifetime, \deltamd is the \BzBzb\
mixing frequency, and $S$ = \slong\ or \stran\ and $C$ = \clong\ or
\ctran\ are the \CP-asymmetry parameters for the longitudinally and
transversely polarized signal. The parameters $S$ and $C$ describe 
\B-mixing induced and direct \CP violation, respectively. 
$S$ and $C$ for the longitudinally
polarized wrong-track signal are fixed to zero. The \deltat PDF
 takes into account incorrect tags and is convolved with
the resolution function described below. Since \ptrue\ is approximately $1$, the fit has no
sensitivity to either \stran\ or \ctran. We set these parameters
to zero and vary them in the evaluation of systematic
uncertainties.

The signal \deltat\ resolution function consists of three
Gaussians ($\sim$$90\%$ core, $\sim$9$\%$ tail, $\sim$1$\%$ outliers),
and takes into account the per-event error on $\deltat$ from the 
vertex fit. The resolution is parameterized using a large sample of fully
reconstructed hadronic \B decays~\cite{prdsin2b}. For wrong-track 
SCF we replace the \B-meson lifetime by an effective lifetime obtained from 
MC simulation to account for the difference in the resolution. The nominal \deltat
distribution for the \B backgrounds is a NP
representation of the MC samples; in the study of systematic
errors we replace this model with the one used for signal. The
resolution for continuum background is described by the sum of three Gaussian
distributions whose parameters are determined from data.

\begin{figure}[!tp]
\begin{center}
\resizebox{8.5cm}{!}{
 \includegraphics{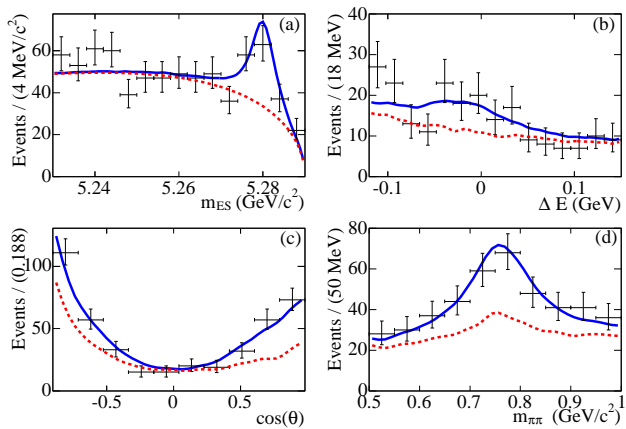}
}
\caption{The distributions for the highest purity tagged events for
the variables \mes (a),  \DeltaE (b),  cosine of the $\rho$ helicity
angle (c) and  \mv (d).  The dotted lines are the sum of backgrounds and the 
solid lines are the full PDF.
} \label{fig:plots}
\end{center}
\end{figure}

\begin{figure}[!t]
\begin{center}
\resizebox{8cm}{!}{
 \includegraphics{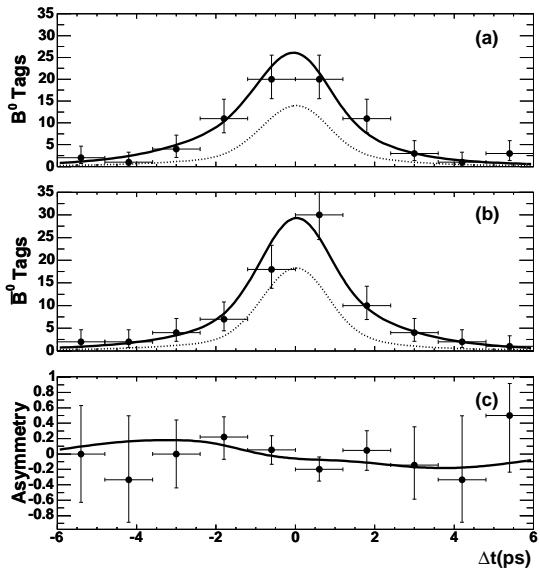}
}
 \caption{The \deltat\ distribution for a sample of events enriched in signal for $\Bz$ (a) and $\Bzb$ (b) tagged events. The dotted lines are the sum of backgrounds and the solid lines are the sum of signal and backgrounds.  The time-dependent \CP asymmetry (see text) is shown in (c), where the curve is the measured asymmetry.} \label{fig:dtplots}
\end{center}
\end{figure}

We perform an unbinned extended maximum likelihood fit. The results of the
fit are $617 \pm 52$ signal events, after correction of a $68$ 
event fit bias, with $\fL = 0.978 \pm 0.014$,
$\slong = -0.33 \pm 0.24$ and $\clong = -0.03 \pm 0.18$. 
The measured signal yield, polarization, and
\CP parameters are in agreement with our earlier
publication~\cite{ref:us}, with significantly improved precision.
Figure~\ref{fig:plots} shows
distributions of \mes, \DeltaE, \coshel\ and \mv\ for the highest purity 
tagged events with a loose 
requirement on \nno. The plot of \mes\ contains 14\% of the 
signal and 1.5\% of the background. For the other plots 
there is an added constraint that $\mes > 5.27 \gevcc$; these 
requirements retain 11.5\% of the signal and 0.4\% of the background.  
Figure~\ref{fig:dtplots} shows
the \deltat\ distribution for \Bz and \Bzb
tagged events. The time-dependent decay-rate asymmetry 
$[  N (\deltat) - \overline{N}(\deltat) ] / [  N (\deltat) + \overline{N}(\deltat) ]$
is also shown, where $N$ $(\overline{N})$ is the decay-rate for \Bz(\Bzb) tagged events.

We have studied possible sources of systematic uncertainties on 
\ptrue, \slong\ and \clong.
The dominant 
uncertainties for \ptrue\ come from
floating the \B background yields ($\,^{+0.00}_{-0.02}$), 
non-resonant events (0.015) and fit bias (0.01). 
The dominant systematic uncertainty on the \CP results comes from the uncertainty
in the \B-background branching ratios. This results in a shift on \slong\ (\clong), 
as large as $\,^{+0.00}_{-0.12}$ ($\,^{+0.008}_{-0.003}$).  
Additional uncertainties on the \CP results come from possible
\CP violation in the \B background, calculated as in Ref.~\cite{ref:us}. 
We allow for a \CP asymmetry up to 20\% in \B\ decays
to final states with charm, resulting in an uncertainty of 0.027
(0.045) on \slong\ (\clong). Allowing for
possible \CP violation in the transverse polarization results in an uncertainty of
0.02 ($\,^{+0.002}_{-0.016}$) on \slong\ (\clong). We estimate
the systematic error on our \CP\ results 
from neglecting
the interference between \Bztorhoprhom and other $4\pi$ final
states: $\B \to a_1\pi$, $\rho \pi\pi^0$ and $\B\to\pi\pi\piz\piz$. 
Strong phases and \CP\ content of the interfering states
are varied between zero and maximum using uniform prior distributions,
 and the RMS deviation of the parameters
from nominal is taken as the systematic error; this is found to be 0.02 
on \slong\ and \clong. Other contributions that are large include
knowledge of the vertex detector alignment 0.034 (0.005) on
\slong\ (\clong), and possible \CP violation in the
doubly-Cabibbo-suppressed decays on the tag side of the
event~\cite{ref:dcsd}.  We allow \CP violation in the wrong-track
SCF to vary between $-1$ and $+1$, which results in changes of
0.007 (0.012) in \slong\ (\clong). The nominal fit does not
account for non-resonant background. If we add a non-resonant
component of $\B\to\rho \pi\pi^0$ events to the likelihood, we fit
$83 \pm 59$ non-resonant events and observe only a $(6 \pm 4)$\% drop in
signal yield. This effect is included in our total systematic uncertainty.
Possible contributions from $\sigma(400) \pi^0\pi^0$ decays are neglected 
due to the small reconstruction efficiency ($0.4\%$).  
Our results are \vspace{-0.9cm}
\begin{center}
\begin{eqnarray}
\ptrue   &=& 0.978 \pm 0.014 \stat \,^{+0.021}_{-0.029} \syst, \nonumber\\
{\slong} &=& -0.33 \pm 0.24 \stat ^{+0.08}_{-0.14} \syst, \nonumber\\
\clong   &=& -0.03 \pm 0.18 \stat \pm 0.09 \syst, \nonumber
\end{eqnarray}
\end{center}\vspace{-0.2cm}
where the correlation between {\slong} and \clong\ is
$-0.042$. 

We constrain the CKM angle $\alpha$ from an isospin analysis
~\cite{grossmanquinn} of $B \to \rho\rho$.  The inputs to the
isospin analysis are the amplitudes of the \CP-even longitudinal
polarization of the $\rho\rho$ final state,
 as well as the measured values 
of \slong\ and \clong\
for \Bztorhoprhom.  We use the measurements of \ptrue, \slong\
and \clong\ presented here; the branching fraction of
$\B^0\to\rhop\rhom$ from ~\cite{ref:us}, 
which uses information from \cite{ref:rhorhoprd}; 
the combined branching fraction and \ptrue\ for $\B\to\rhop\rhoz$
from Ref.~\cite{recentrhorho}; the central value corresponding to the upper 
limit of ${ \cal B}(\B\to\rhoz\rhoz)$ from Ref.~\cite{PRLrho0rho0}. We
ignore electroweak penguins and possible $I=1$ amplitudes
\cite{falk}.

\begin{figure}[!ht]
 \begin{center}
 \includegraphics[width=0.48\textwidth]{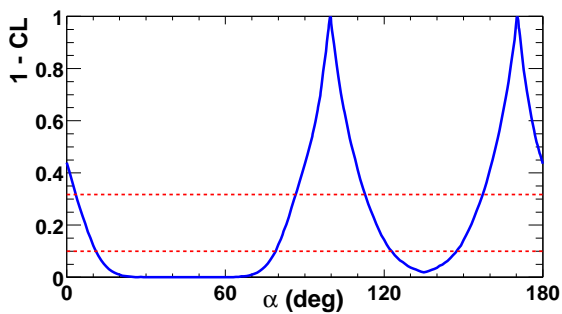}
\caption{ CL on $\alpha$ obtained from the isospin analysis 
with the statistical method described in~\cite{ref:ckmbestfit}. 
The dashed lines correspond to the 68\% (top) and 90\% (bottom) 
CL intervals. } \label{fig:alpha}
 \end{center}
\end{figure}

To interpret our results in terms of a constraint on $\alpha$ from
the isospin relations,  we construct a $\chi^2$ that includes
the measured quantities expressed as the lengths of the sides of
the isospin triangles and we determine the minimum $\chi^2_0$.
As the isospin triangles do not close with the current central 
values of the branching ratios, we have adopted a toy MC
techniques to compute the confidence level (CL) on $\alpha$; our method
is similar to
the approach proposed in Ref.~\cite{FC98}. 
For each value of $\alpha$, scanned between
$0$ and $180^\circ$, we determine the
difference $\Delta \chi^2_{{\rm DATA}}(\alpha)$  between  the minimum
of $\chi^2(\alpha)$ and $\chi^2_0$. We then generate MC experiments
around the central values obtained from the fit to data with
the given value of $\alpha$ and
we apply the same procedure. The fraction of these experiments in
which  $\Delta \chi^2_{{\rm MC}}(\alpha)$ is smaller than
$\Delta \chi^2_{{\rm DATA}}(\alpha)$ 
is interpreted as the CL on $\alpha$.
Figure~\ref{fig:alpha} shows $1-{\rm CL}$ for
$\alpha$ obtained from this method.
Selecting the solution closest to the CKM
combined fit average~\cite{ref:ckmbestfit,ref:utfit} we find $\alpha =
100^\circ\pm13^\circ$, where the error is
dominated by $\delta\alpha_{\rho\rho}$ which is $\pm 11^\circ$ at $1\sigma$. 
The  90\% CL allowed interval for $\alpha$ is between $79^\circ$ and
$123^\circ$.

In summary we have improved the measurement of the \CP-violating parameters
\slong\ and \clong\ in \Bztorhoprhom using a data-sample 2.6 times larger 
than that in Ref.~\cite{ref:us}. We do not observe mixing-induced or direct 
\CP violation. We derive a model-independent measurement of the 
CKM angle $\alpha$, which is the most precise to date.

We are grateful for the excellent luminosity and machine conditions
provided by our \pep2\ colleagues, 
and for the substantial dedicated effort from
the computing organizations that support \babar.
The collaborating institutions wish to thank 
SLAC for its support and kind hospitality. 
This work is supported by
DOE
and NSF (USA),
NSERC (Canada),
IHEP (China),
CEA and
CNRS-IN2P3
(France),
BMBF and DFG
(Germany),
INFN (Italy),
FOM (The Netherlands),
NFR (Norway),
MIST (Russia), and
PPARC (United Kingdom). 
Individuals have received support from CONACyT (Mexico), A.~P.~Sloan Foundation, 
Research Corporation,
and Alexander von Humboldt Foundation.

\end{document}